\documentclass{PoS}
\def\aw{Alfv\'en}
\def\gp{$\Gamma^\prime$}

\title{Non-linear Cosmic Ray propagation close to the acceleration site}

\ShortTitle{Non linear CR propagation}

\author{\speaker{Lara Nava}
\\
        The Hebrew University of Jerusalem, Jerusalem, 91904, Israel\\
        E-mail: \email{lara.nava@mail.huji.ac.il}}

\author{Stefano Gabici\\
        Astroparticule et Cosmologie (APC), CNRS, Université Paris 7 Denis Diderot, Paris, France\\
        E-mail: \email{stefano.gabici@apc.univ-paris7.fr}}

\author{Alexandre Marcowith\\
Laboratoire Univers et particules de Montpellier, Université Montpellier/CNRS, 34095 Montpellier, France\\
E-mail: \email{alexandre.marcowith@umontpellier.fr}}

\author{Giovanni Morlino\\
INFN -- Gran Sasso Science Institute, viale F. Crispi 7, 67100 L'Aquila, Italy\\
E-mail: \email{giovanni.morlino@gssi.infn.it}}

\author{Vladimir Ptuskin\\
Pushkov Institute of Terrestrial Magnetism, Ionosphere and Radio Wave Propagation of the Russian Academy of Sciences, Troitsk, Moscow, 142190 Russia\\
E-mail: \email{vptuskin@izmiran.ru}}

\abstract{Recent advances on gamma-ray observations from SuperNova Remnants and Molecular Clouds offer the possibility to study in detail the properties of the propagation of escaping Cosmic Rays (CR). However, a complete theory for CR transport outside the acceleration site has not been developed yet. Two physical processes are thought to be relevant to regulate the transport: the growth of waves caused by streaming instability, and possible wave damping mechanisms that reduce the growth of the turbulence. Only a few attempts have been made so far to incorporate these mechanisms in the theory of CR diffusion. In this work we present recent advances in this subject. In particular, we show results obtained by solving the coupled equations for the diffusion of CRs and the evolution of \aw\ waves. We discuss the importance of streaming instabilities and wave damping in different ISM phases.}

\FullConference{The 34th International Cosmic Ray Conference,\\
		30 July- 6 August, 2015\\
		The Hague, The Netherlands}

\begin{document}
\section{Introduction}
\label{sect:introduction}
We study the non-linear diffusion of a population of Cosmic Rays (CRs) after they escape from the acceleration site (for example, a supernova remnant shock). As a zero order approximation the problem can be described by an impulsive injection of CRs of a given energy at a given location in the Galaxy. 
We study the transport of CRs taking into account the growth of \aw\ waves caused by particle-wave interactions, and their back-reaction on the CR diffusion.
We also consider possible linear wave damping mechanisms that limit the growth of the waves. The problem is described by two differential equations that govern the evolution of the CR pressure and of the wave energy density as a function of time and distance from the acceleration site. We present numerical solutions of these two coupled equations for different values of the physical parameters involved in the problem.

\section{The model}
\label{sect:model}
In our model the transport of CRs is assumed to be regulated by the resonant scattering off \aw\ waves, i.e. a CR of energy $E$ resonates with waves of wave number $k = 1/r_L(E)$ where $r_L$ is the particle Larmor radius (see e.g. \cite{wentzel}). The (normalized) energy density $I(k)$ of \aw\ waves is defined as:
\begin{equation}
\frac{\delta B^2}{8 \pi} = \frac{B_0^2}{8 \pi} \int I(k) {\rm d}\ln k ~ ,
\end{equation}
where $B_0$ is the ambient magnetic field and $\delta B$ the amplitude of the magnetic field fluctuations.

According to quasi--linear theory, CRs diffuse along the magnetic field lines with a diffusion coefficient equal to \cite{bell78}:
\begin{equation}
\label{eq:diff}
D = \frac{4 ~ c ~ r_L(E)}{3 \pi~ I(k)} = \frac{D_B(E)}{I(k)} ~ ,
\end{equation}
which can be expressed as the ratio between the Bohm diffusion coefficient $D_B(E)$ and the energy density of resonant waves $I(k = 1/r_L)$. Formally, quasi--linear theory is valid for $\delta B/B_0 \ll 1$. In this limit the diffusion perpendicular to the field lines can be neglected, being suppressed by a factor of $(\delta B_k/B_0)^4 \equiv I(k)^2$ with respect to the one parallel to $B_0$ (see e.g. \cite{drury1983,fabien}). This implies that under the condition that $\delta B/B_0$ remains small, the problem is one--dimensional.

We consider the streaming of CR to be the main source of Alfv\'enic turbulence. 
For CRs that stream along the direction of $B_0$ (that we consider aligned along the coordinate $z$) the growth rate of \aw\ waves $\Gamma_{CR}$ is proportional to the product between the \aw\ speed $V_A$ and the gradient of the pressure of resonant CRs, and can be defined as (see e.g. \cite{drury1983,skilling1975}):
\begin{equation}
\label{eq:growth}
2\Gamma_{CR} I = - V_A \frac{\partial P_{CR}}{\partial z} ~ .
\end{equation}
The sign indicates that only waves traveling in the direction of the streaming are excited. For dimensional consistency, the pressure of CRs $P_{CR}$ has to be normalized to the magnetic energy density $B_0^2/8\pi$.

On scales smaller than the magnetic field coherence length, a flux tube characterized by a constant magnetic field strength $B_0$ and directed along the $z$--axis can be considered.
Two coupled equations (for the evolution of CRs and waves along the flux tube) must be solved: the diffusion coefficient of CRs of energy $E$ (Eq.~\ref{eq:diff}), indeed, depends on the energy density of resonant waves $I(k)$, and in turn the growth rate of these waves (Eq.~\ref{eq:growth}) depends on the gradient of the pressure of resonant CRs. The two coupled equations then read:
\begin{equation}
\label{eq:CRs}
\frac{\partial P_{CR}}{\partial t} + V_A \frac{\partial P_{CR}}{\partial z} = \frac{\partial}{\partial z} \left( \frac{D_B}{I} \frac{\partial P_{CR}}{\partial z} \right)
\end{equation}
\begin{equation}
\label{eq:waves}
\frac{\partial I}{\partial t} + V_A \frac{\partial I}{\partial z} = - V_A \frac{\partial P_{CR}}{\partial z} - 2\Gamma_d I + Q
\end{equation}
where the left side in both expressions is the time derivative computed along the characteristic of excited waves:
\begin{equation}
\frac{\rm d}{{\rm} d t} = \frac{\partial}{\partial t} + V_A \frac{\partial}{\partial z} ~~ .
\end{equation}
The advective terms $V_A \partial P_{CR}/\partial z$ and $V_A \partial I/\partial z$ are neglected in the following since we verified that they play little role in the situation under examination.
The last two terms in Eq.~\ref{eq:waves} account for possible mechanisms of wave damping, operating at a rate $\Gamma_d$, and for the injection $Q$ of turbulence from an external source (i.e. other than CR streaming). In this work we consider only linear damping mechanisms, for which the damping rate is independent of space and time. The term representing the external source of turbulence can be set to $Q =2 \Gamma_d I_0$, so that when streaming instability is not relevant the level of the background turbulence is at a constant level $I = I_0$.  

As already pointed out by \cite{malkov}, the approach described above decouples the process of acceleration of particles, which operates, for example, at a SNR shock, from the particle escape from the acceleration site. Though such a separation might seem artificial, the problem defined above can still be useful to describe the escape of particles form a {\it dead} accelerator, in which the acceleration mechanism does not operate any more, or operates at a much reduced efficiency \cite{malkov}. This situation would probably apply to the case of old SNRs.
On the other hand, \cite{plesser} suggested that Equations.~\ref{eq:CRs} and \ref{eq:waves} could be also used to describe an intermediate phase of CR propagation in which the CRs have left the source but are not yet completely mixed with the CR background. For the case of a SNR shock, the equations above could thus be applied to those CRs characterized by a diffusion length large enough to decouple them from the shock region. Typically, this happens during the Sedov phase to the highest energy particles confined at the SNR shock when the diffusion length $D_B/u_s$ gradually increases with time  up to a value larger than some fraction $\chi$ of the SNR shock radius $R_s$, where $u_s$ is the shock speed and $\chi \approx 0.05 ... 0.1$ \cite{ptuskinzirakashvili2005,gabiciescape}.

In both scenarios, the initial conditions for Eqs~\ref{eq:CRs} and \ref{eq:waves} can be set as follows:
\begin{eqnarray}
P_{CR} &=& P_{CR}^0 ~~~~~~~ {\rm and} ~~~~~~~~  I \approx 1 ~~~~~~~~~~~~~~~~ {\rm for} ~~~~~~~ z < a \\
P_{CR} &=&  0 ~~~~~~~~~~~{\rm and} ~~~~~~~~      I  =  I_0 \ll 1 ~~~~~~~{\rm for} ~~~~~~~ z > a
\end{eqnarray}
where $a$ is the spatial scale of the region filled by CRs at the time of their escape from the source.

Following \cite{malkov}, we introduce the quantity $\Pi$, defined as:
\begin{equation}
\Pi ~=~ \frac{V_A}{D_B} ~ \Phi_{CR}
\label{eq:pai}
\end{equation}
where:
\begin{equation}
\Phi_{CR} \equiv \int_0^{\infty} {\rm d}z ~ P_{CR} = P_{CR}^0 a ~ .
\end{equation}
To understand the physical meaning of $\Pi$, consider the initial setup of the problem, in which CRs are localized in a region of size $a$. The CR pressure within $a$ is $P_{CR}^0$, and then $\Phi_{CR} = P_{CR}^0 a$. The initial diffusion coefficient outside the region of size $a$ is equal to $D_B/I_0$. 
The problem of transport is characterised by three relevant timescales: i) the growth time $\tau_g \approx (V_A/I_0 \partial P_{CR} /\partial z)^{-1} \approx a I_0/V_A P_{CR}^0$, ii) the time it takes the CR cloud to spread due to diffusion $\tau_{diff} \approx a^2 / D \approx a^2 I_0/D_B$, and iii) the damping time $\tau_d = 1/2\Gamma_d$. 
In order to have a significant growth of waves due to CR streaming, the timescale for growth must be shorter than the two other timescales:
$\tau_g<\min(\tau_{diff},\tau_d)$. In terms of the parameter $\Pi$ this conditions reads: $\Pi > \max(1,\tau_{diff}/\tau_d)$.
It is evident, then, that the parameter $\Pi$ regulates the effectiveness of CR--induced growth of waves. In the absence of a damping term for waves, $\Pi > 1$ is a necessary condition for streaming instability to be relevant, while in the presence of efficient wave damping, a more stringent condition on $\Pi$ applies.

For $\Pi \lesssim \max(1,\tau_{diff}/\tau_d)$ CRs play no role in the generation of \aw\ waves, and Equation~\ref{eq:CRs} can be solved analytically:
\begin{equation}
\label{eq:test}
P_{CR} = \sqrt{\frac{I_0}{\pi D_B t}} ~ \Phi_{CR} ~ e^{-\frac{I_0 z^2}{4 D_B t}}
\end{equation}
Equation~\ref{eq:test} is referred to as {\it test--particle solution}.
When wave growth cannot be neglected, the solution deviates from this analytic test-particle solution.
In the extreme scenario $\Pi >> \max(1,\tau_{diff}/\tau_d)$ waves grow so quickly that the the diffusive term in Eq.~\ref{eq:CRs} becomes negligible when compared to advection term $V_A \partial P_{CR}/\partial z$, and the advection term cannot be neglected anymore. This describes a situation in which CRs are "locked" to waves and move with them at a velocity equal to $V_A$ \cite{skilling1971}. An identical result was found by \cite{cesarsky1975} in a study of the escape of $\approx$~MeV CRs from sources, and also suggested by \cite{hartquist}.

\section{The method}
\label{sect:method}
\begin{figure}
{\includegraphics[scale=0.38]{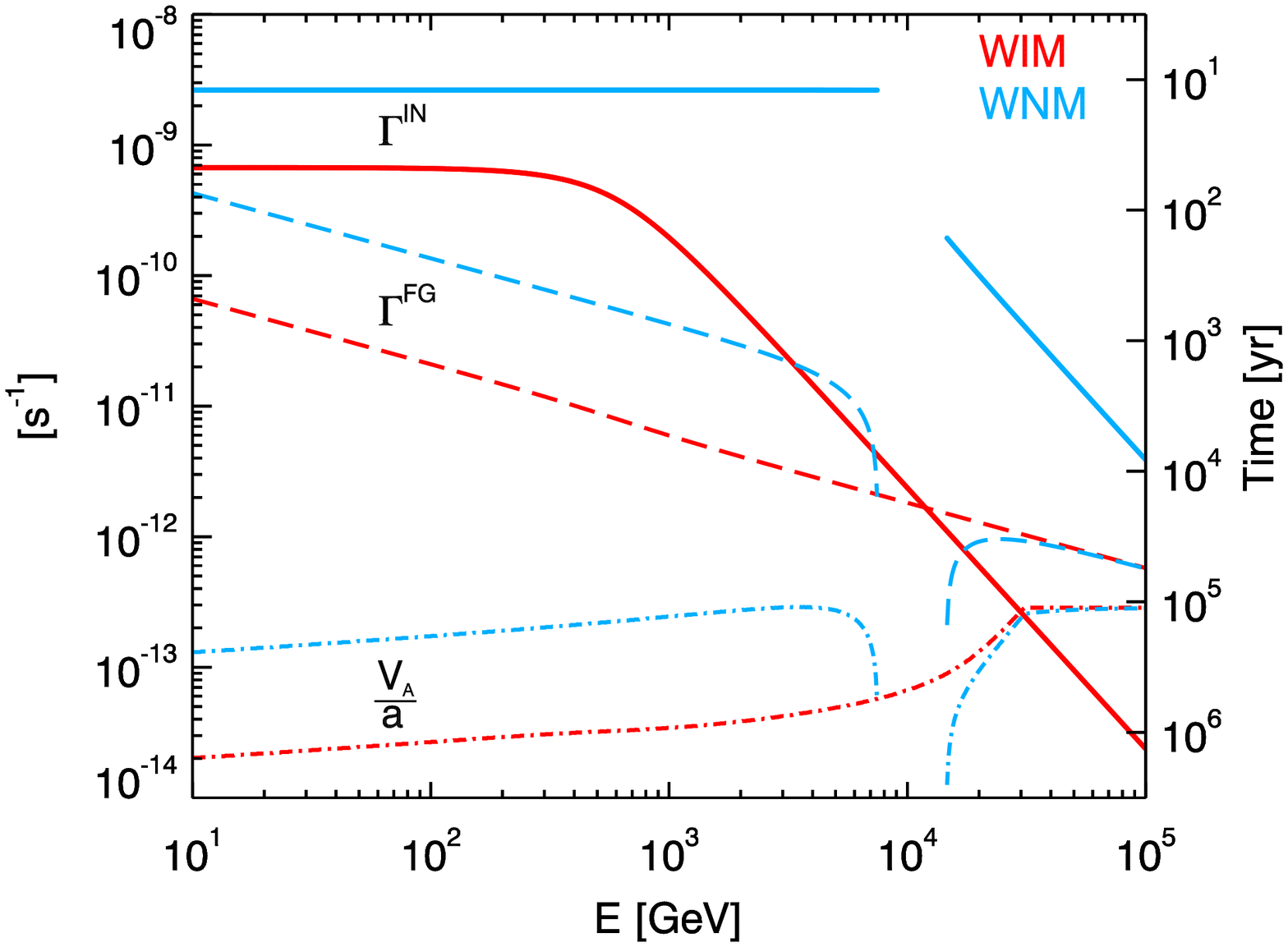}
\includegraphics[scale=0.38]{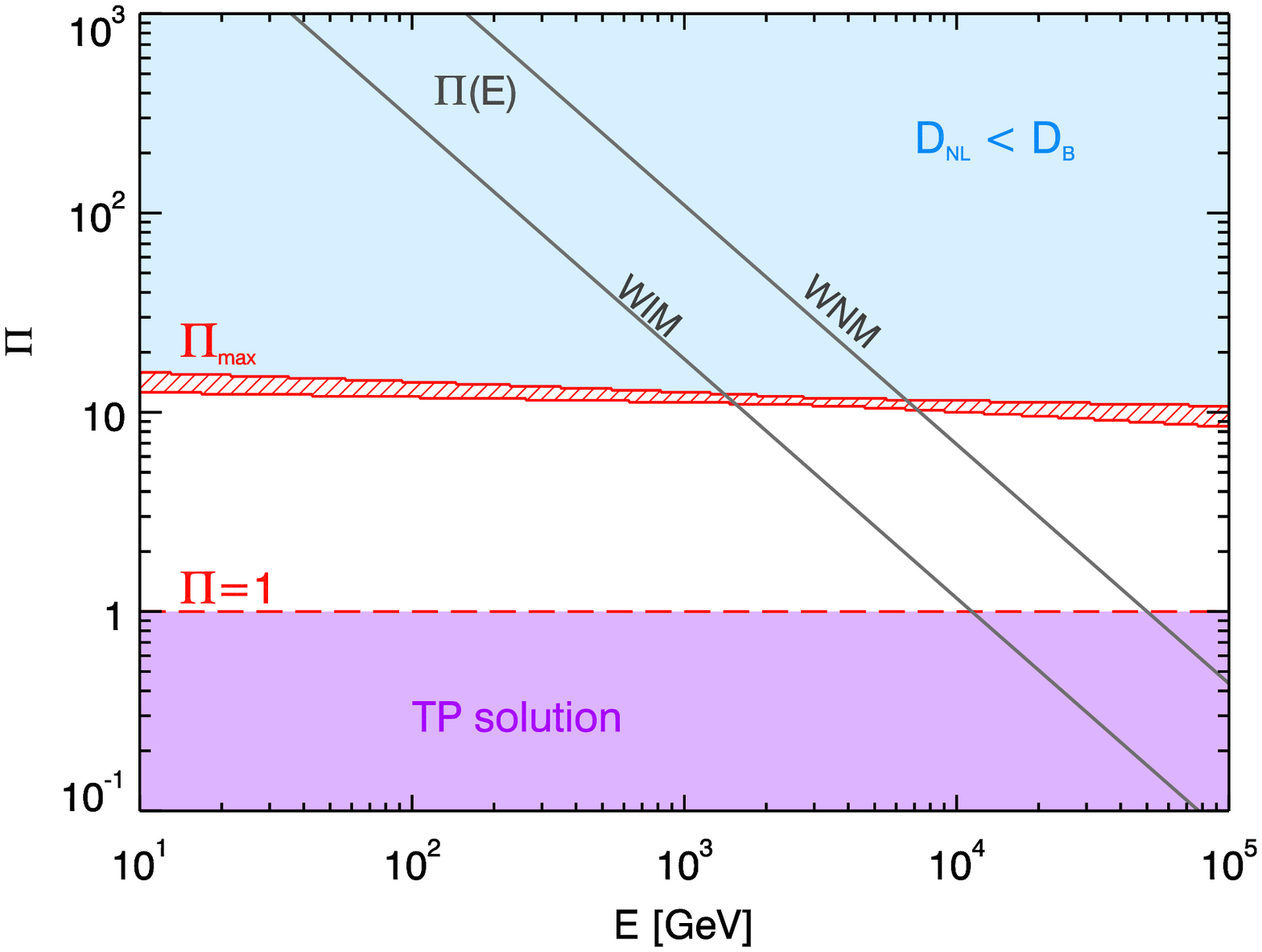}}
\caption{Left: ion-neutral damping rate ($\Gamma^{IN}_d$, solid curves), Farmer \& Goldreich damping rate ($\Gamma^{FG}_d$, dashed curves) and ratio between \aw\ speed and initial size of the source ($V_A/a$, dot-dashed curves), as a function of the CR energy $E$. Different colours refer to two different ISM phases: warm ionised medium (WIM, red), and warm neutral medium (WNM, blue).
Right: parameter $\Pi$ as a function of the CR energy for a WIM and WNM (solid lines). In absence of efficient wave damping,  values of $\Pi<1$ correspond to a test particle solutions, since the wave growth rate is inefficient. Conversely, for $\Pi$ larger that $\Pi_{max}\sim10$ the growth rate is large and the diffusion coefficient becomes smaller than the Bohm diffusion coefficient $D_B$. In presence of efficient wave damping, both these limits are shifted to higher values, depending on the value of $\Gamma_d$.}
\label{fig:damp}
\end{figure}
In order to solve Eqns.~\ref{eq:CRs} and \ref{eq:waves} it is convenient to perform the following change of coordinates \cite{malkov}:
\begin{equation}
s\equiv\frac{z}{a}  \qquad\qquad  
\tau\equiv \left(\frac{V_A}{a}\right) t 
\end{equation}
and re-normalise the parameters as follow:
\begin{equation}
\mathcal{P}_{CR}\equiv\left( \frac{V_A a}{D_B} \right) P_{CR} \qquad\qquad
W\equiv\left( \frac{V_A a}{D_B} \right) I  \qquad\qquad
\Gamma^\prime\equiv\left( \frac{a}{V_A} \right) \Gamma_d
\end{equation}
In these notations (and after dropping the advection terms) the equations become:
\begin{eqnarray}
\frac{\partial \mathcal{P}_{CR}}{\partial \tau} &=& \frac{\partial}{\partial s} \left( \frac{1}{W} \frac{\partial \mathcal{P}_{CR}}{\partial s} \right) \label{eq:norm_p}\\
\frac{\partial W}{\partial \tau} &=& - \frac{\partial \mathcal{P}_{CR}}{\partial s} - 2\Gamma^\prime \left( W - W_0 \right) \label{eq:norm_w}
\end{eqnarray}

Their solution (i.e., $\mathcal{P}$ and $W$ as a function of the variables $\tau$ and $s$) depends only on three parameters: the initial values $\mathcal{P}_{CR}^0$, $W_0$, and $\Gamma^\prime_0$. Note that: i) $\mathcal{P}_{CR}^0=\Pi$, ii) $W_0=V_Aa/D_0$, and iii) $\Gamma^\prime=\Gamma^\prime_0$ (since we consider here only a linear damping constant in both time and space).
In these notations, the condition for effective growth of waves is $\Pi > \max(1,\Gamma^\prime W_0)$.

We present examples of numerical solutions to the coupled equations \ref{eq:norm_p} and \ref{eq:norm_w} for values of the parameters in the ranges: $\Pi=[2-20]$, $W_0=[10^{-5}-10^{-2}]$, and $\Gamma^\prime=[0-100]$. 
In order to understand why we chose these values, and to which physical conditions they correspond, it is necessary to specify the values of $V_A, a, D_0$, $D_B$, and the damping mechanisms. These quantities in general depend on the properties of the ISM, and on the particle energy. We consider two different ISM phases: a warm neutral medium (WNM) and a warm ionised medium (WIM). The typical values of density, temperature, ionisation fraction, and magnetic field strength for these two phases are taken from \cite{jean}.
In these ISM phases, the two most relevant linear damping mechanisms are the ion-neutral friction ($\Gamma_d^{IN}$) and the damping by background MHD turbulence suggested by Farmer \& Goldreich  \cite{fg} ($\Gamma_d^{FG}$). The value of $\Gamma_d^{FG}$ as a function of the particle energy is shown in Fig.~\ref{fig:damp} (left panel, long-dashed lines), for both ISM phases.
The ion-neutral damping rate is estimated by numerically solving the equations in \cite{zweibel}. The results are shown in Fig.~\ref{fig:damp} (left panel, solid lines), for both ISM phases: at low energies $\Gamma^{IN}_d=\nu_{IN}/2$ (where $\nu_{IN}$ is the ion-neutral collision frequency), while at high energies $\Gamma^{IN}_d\propto E^{-2}$. The level of coupling between ions and neutrals affects also the \aw\ speed, that is in the range $10^6-10^7\,$cm/s.
Since it is useful for our calculations, we show in Fig.~\ref{fig:damp} the value of the ratio $V_a/a$, as a function of energy (dash-dotted lines).
In order to estimate $a$ we consider $a\approx R_{esc}(E)$, where $R_{esc}(E)$ is the escape radius and depends on the CR energy. We model its dependence on the energy following \cite{cristofari}: we found that its value ranges between a few pc (for the highest energy particles) and $\sim$20 pc (for $\sim$GeV particles). 
Since we are interested in following the CR diffusion from the scale $a$ where they are released up to a distance $z\lesssim10^3\,$pc, this roughly corresponds to solve the normalised equations in a range of $s$ from 1 to several hundreds.
Moreover, we want to study the CR pressure at different times $t$ in the range $10^3\,$yr to $\lesssim 10^5\,$yr. Considering the value of $V_a/a$ as a function of energy displayed in Fig.~\ref{fig:damp}, this corresponds to $\tau= 10^{-2}-1$. 

In order to identify the relevant values of the parameters $\Pi, W_0, \Gamma^\prime$ we proceed as follow:
\begin{itemize}
\item \gp -- the damping rate $\Gamma_d=max(\Gamma^{IN}_d,\Gamma^{FG}_d)$ ranges between $10^{-9}$ and  $10^{-12}\,$s$^{-1}$ (Fig.~\ref{fig:damp}, left panel), depending on the particle energy and ISM phase. This corresponds to $1<\Gamma^\prime\lesssim10^4$.

\item $\Pi$ -- the values of $\Pi$ as a function of the CR energy, for the two different ISM is shown in Fig.~\ref{fig:damp} (right panel, solid lines).

\item $W_0$ -- for the background diffusion $D_0$ we can refer to the average Galactic value $D_0=D_{ISM}=10^{28}(E/10\,\rm GeV)^{0.5}\,$cm$^2$/s. Since $V_Aa$ ranges between $10^{27}\,$cm$^2$/s and $10^{25}\,$cm$^2$/s (at low and high energies, respectively), the range of relevant values for $W_0$ is $10^{-5}<W_0<10^{-1}$.

\end{itemize}

\begin{figure}
\hskip -0.55 truecm
\includegraphics[scale=0.89]{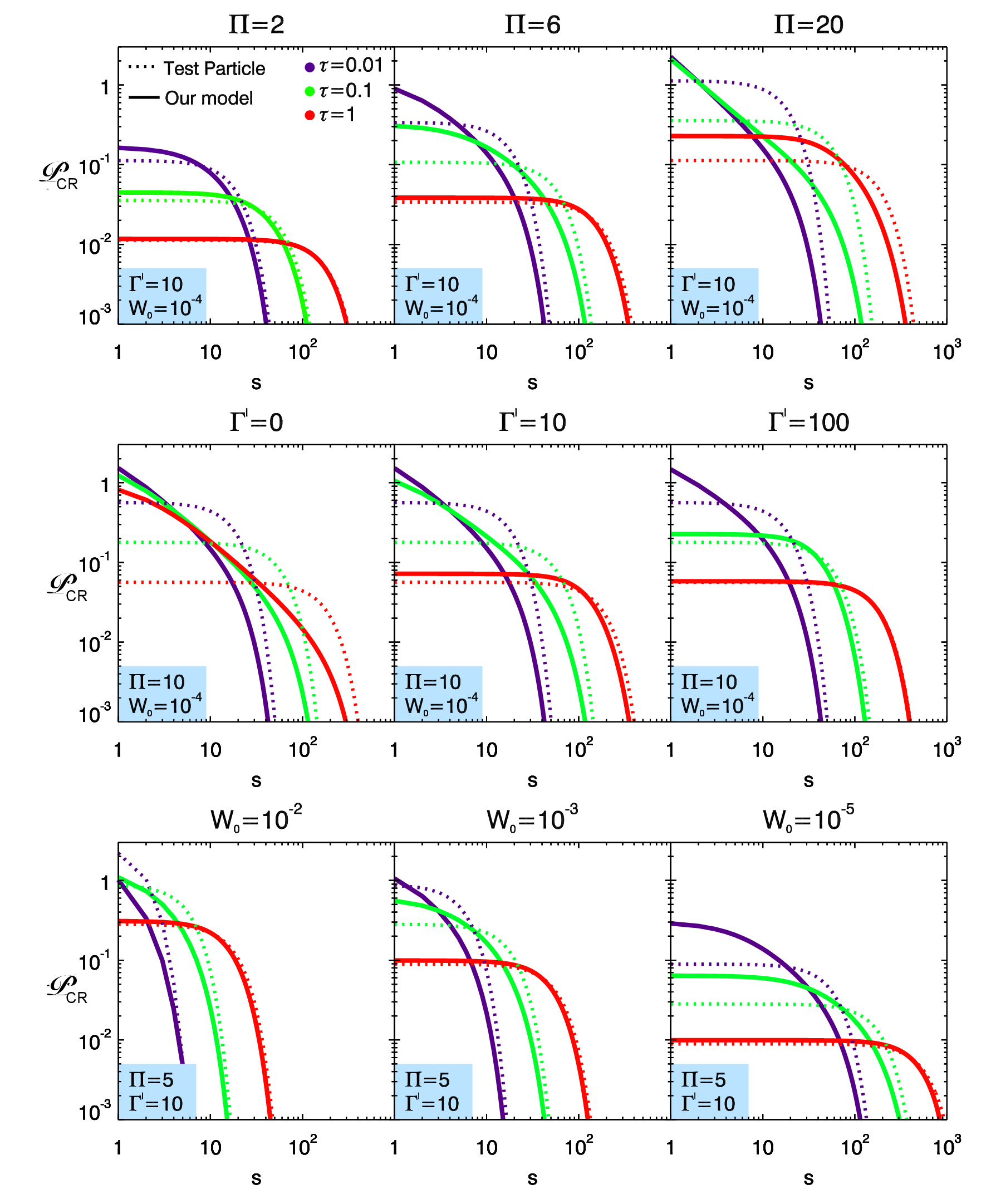}
\caption{Normalized CR pressure as a function of the variable $s$, which represents the distance from the source $z$ normalised to the size of the source $a$. In each row the results are shown by varying only one of the three parameters (whose value is reported above each panel) and keeping the other two fixed to the values reported in the shaded area. The solid curves show the results from the numerical computation, while dotted curves represent the test particle solution, where the role of streaming instabilities in enhancing the turbulence is neglected. Different colours refer to different normalised times $\tau=10^{-2}, 10^{-1},1$.}
\label{fig:main}
\end{figure}

\section{Results and Discussion}
\label{sect:results}
We numerically solve Eqs.~\ref{eq:norm_p} and \ref{eq:norm_w} and show the results for $\mathcal{P}(s,\tau)$ vs. $s$ (solid lines in Fig.~\ref{fig:main})
at three different normalised times $\tau=10^{-2},10^{-1},1$ (corresponding to different colours in the figure).
The test particle solution is also shown for comparison (dotted lines).
Each row in the figure corresponds to results obtained by varying only one parameter at the time. 

The upper row shows the impact of the parameter $\Pi$ for fixed values of $\Gamma^\prime=10$ and $W_0=10^{-4}$. Since for $\Pi<1$ the numerical solution coincides with the test particle solution, we show our results for values of $\Pi>1$. When $\Pi=2$, the solution starts to deviate from the test particle case: due to the growth of wave by streaming instability the diffusion is slower, the CR pressure in the vicinity of the accelerator is larger, and the diffusion length is smaller. The effect is more and more evident for increasing values of $\Pi$ (upper row, middle and right panels).

The middle row shows the effect of wave damping, for fixed $\Pi=10$ and $W_0=10^{-4}$. Since $\Pi>>1$, when wave damping is negligible (\gp=0, left panel) the diffusion is suppressed as compared to the test particle case. Larger values of \gp\ limit the growth of \aw\ waves and the solution approaches the test particle solution. Note that \gp=10 (middle row, central panel) corresponds to a damping mechanism that becomes important at $\tau>0.1$. For this reason the solution at $\tau=0.01$ and $\tau=0.1$ are unaffected by the increased damping rate, and only the solution at late time $\tau=1$ is modified, and approaches the test particle solution. The right panel shows the case of a even faster damping mechanism, \gp=100, whose effect starts to be relevant at times corresponding to $\tau>0.01$, explaining why the solution at intermediate and late times approaches the test particle solution, while the solution at early time is unaffected by wave damping.

The bottom row shows calculations performed at $\Pi=5$ and \gp=10, with $W_0$ varying from $10^{-5}$ to $10^{-2}$.
A large $W_0$ (left panel) corresponds to slow diffusion and small diffusion length, both in the test particle case and in the numerical solution. The two solutions differ at early and intermediate times because the growth rate is faster than the diffusion rate, and suppresses the diffusion. At later times instead the growth rate decreases and the numerical solution does not appreciably differ from the test particle case. A change in the initial value of the background turbulence has the same effect on both the analytical and numerical solution: the diffusion is faster for decreasing values of $W_0$ (see bottom row, from left to right), and the CRs can reach larger distances. The deviation from a test particle solution however remains unchanged, since both the growth rate and the diffusion rate scale as $W_0$.

This study outlines the importance of considering, in the physics of CR transport, both the growth of waves due to streaming instability and the role of mechanisms that damp the turbulence. When damping is neglected, even relatively small values of the integrated CR pressure $\Pi=10$ (corresponding, for example, to particles of a few TeV, in a WIM - see Fig.~\ref{fig:damp}) in a environment with small background turbulence $W_0=10^{-4}$ (relevant for TeV particles when the background diffusion coefficient $D_0$ is similar to the average Galactic diffusion coefficient) lead to the conclusion that diffusion is strongly suppressed by the efficient growth of waves, and particles cannot travel large distances, even at late times. This situation is shown in Fig.~\ref{fig:main} (middle row, left panel) where the numerical computation is performed in absence of a wave damping term: the diffusion is strongly suppressed also at $\tau=1$, corresponding to old remnants ($t_{age}\sim10^5\,$yr).

\end{document}